\documentclass[twocolumn,aps,prc,superscriptaddress,showpacs,floatfix,longbibliography,nofootinbib]{revtex4-1}
\usepackage{url}
\usepackage{cancel}
\usepackage[colorlinks,linkcolor=blue,citecolor=blue,filecolor=black,urlcolor=blue]{hyperref}
\usepackage{epsfig,graphics}
\usepackage{graphicx}% Include figure files
\usepackage{dcolumn}% Align table columns on decimal point
\usepackage{bm}% bold math
\usepackage[usenames]{color}
\usepackage{amssymb}
\usepackage{amsmath}
\usepackage{multirow}
\usepackage{float}
\usepackage{harpoon}
\usepackage{MnSymbol}
\usepackage{appendix}
\usepackage{color}
\usepackage{hyperref}
\usepackage{cleveref}
\usepackage{dutchcal}
\usepackage[normalem]{ulem}
\begin{document}

\title{Is nucleon spin thermalized in intermediate-energy heavy-ion collisions?}
\author{Jun Xu}\email{junxu@tongji.edu.cn}
\affiliation{School of Physics Science and Engineering, Tongji University, Shanghai 200092, China}
\affiliation{Southern Center for Nuclear-Science Theory, Institute of Modern Physics,
Chinese Academy of Sciences, Guangdong 516008, China}
\begin{abstract}
Despite the success of the spin-thermalized assumption in explaining hyperon spin polarizations in relativistic heavy-ion collisions, challenges begin to arise especially at lower collision energies. The present study compares the nucleon spin polarization during the collision process and at the freeze-out stage from a non-relativistic spin-dependent transport model with spin-thermalized approaches in intermediate-energy heavy-ion collisions, where the relativistic effect and the temperature gradient have shown to be unimportant. It is found that both the global and local spin polarizations are largely overestimated from spin-thermalized approaches, compared to those generated by the spin-orbit mean-field potential in transport simulations.
\end{abstract}
\maketitle

%\section{Introduction}
%\label{introduction}

Spin physics is generally related to various exotic phenomena in many research fields, and the spin-orbit coupling, despite its different forms in various systems~\cite{PhysRevB.92.245425}, is the key source of the spin physics. In non-central relativistic heavy-ion collisions, the produced quark-gluon plasma carries part of the angular momentum from the collision and becomes the most vortical fluid produced on the earth. Due to the spin-orbit coupling, the spins of quarks are polarized perpendicular to the reaction plane. Hadrons produced through hadronization carry the information of quark spins, and the resulting spin polarizations of $\Lambda$~\cite{STAR:2017ckg,PhysRevC.98.014910}, $\Omega$~\cite{PhysRevLett.126.162301}, and $\Xi$~\cite{PhysRevLett.126.162301} hyperons as well as the spin alignments of $\phi$ and $K^\star$ mesons~\cite{STAR:2022fan} have been observed experimentally by measuring the angular distribution of their decays. Besides the above global polarization perpendicular to the reaction plane, scientists have also observed some azimuthal angle dependence of the longitudinal spin polarization~\cite{PhysRevLett.123.132301}, indicating the rich physics of spin dynamics. More recently, appreciable $\Lambda$ spin polarizations have been observed in the STAR fix-target experiment~\cite{PhysRevC.104.L061901} and by HADES Collaboration~\cite{2022137506} at lower collision energies dominated by hadronic dynamics.

In ultra-relativistic heavy-ion collisions dominated by partonic degree of freedom, the spin polarization of hyperons as well as its energy dependence~\cite{Karpenko:2016jyx,Li:2017slc} has been successfully described by the spin-thermalized assumption, i.e., by assuming that the particle spin is completely thermalized in the local vorticity field. However, this spin-thermalized assumption is challenged by a few experimental observables. The traditional kinematic vorticity field leads to the wrong azimuthal angle dependence of the longitudinal spin polarization, i.e., the famous ``sign'' problem~\cite{PhysRevLett.120.012302,PhysRevLett.123.132301}, unless the contribution of the thermal shear is included~\cite{Becattini:2021suc,Fu:2021pok} or non-equilibrium dynamics is taken into account~\cite{PhysRevLett.125.062301,PhysRevResearch.1.033058}. In addition, at lower collision energies the $\Lambda$ spin polarization doesn't seem to increase monotonically with decreasing collision energy~\cite{2022137506}, indicating that the particle spin may not be thermalized in collisions with the dynamics dominated by nucleon degrees of freedom.

Obviously, spin-dependent transport models are highly demanded to understand the non-equilibrium hadronic dynamics. While covariant transport models suitable for collisions at energies of a few AGeV are still lacking, a non-relativistic spin- and isospin-dependent Boltzmann-Uehling-Uhlenbeck (SIBUU) transport model with the nucleon spin degree of freedom explicitly included was initiated in Ref.~\cite{Xu:2012hh}, suitable for describing the non-equilibrium spin dynamics at the beam energy around a few hundred AMeV. A consistent derivation of the spin-dependent equations of motion was given later~\cite{Xia:2016xiw}, and the lattice Hamiltonian framework for the mean-field evolution~\cite{Xia:2019whr} as well as the rigorous angular momentum conservation~\cite{Liu:2023nkm} were further incorporated into SIBUU for studying various spin-related observables including the spin polarization. Based on SIBUU, I will investigate whether the nucleon spin is thermalized in the hot nuclear matter produced in intermediate-energy heavy-ion collisions, thus to understand the relation between the nucleon spin and the vorticity field from the low-energy side.

%\section{Theoretical Framework}
%\label{theory}

%\subsection{Basic framework of SIBUU}

Let's begin with the spin-dependent BUU equation for spin-1/2 particles expressed as~\cite{Ring1980,Smith1989}
\begin{eqnarray}\label{BLE}
\frac{\partial \hat{f}}{\partial t}&+&\frac{i}{\hbar}\left [ \hat{\varepsilon},\hat{f}\right]+\frac{1}{2}\left ( \frac{\partial \hat{\varepsilon}}{\partial \vec{p}}\cdot \frac{\partial \hat{f}}{\partial \vec{r}}+\frac{\partial \hat{f}}{\partial \vec{r}}\cdot \frac{\partial \hat{\varepsilon}}{\partial \vec{p}}\right ) \nonumber\\
&-&\frac{1}{2}\left ( \frac{\partial \hat{\varepsilon}}{\partial
\vec{r}}\cdot \frac{\partial \hat{f}}{\partial
\vec{p}}+\frac{\partial \hat{f}}{\partial \vec{p}}\cdot
\frac{\partial \hat{\varepsilon}}{\partial \vec{r}}\right )=I_c.
\end{eqnarray}
Compared to the traditional BUU equation, both the phase-space distribution $\hat{f}$ and the single-particle energy $\hat{\varepsilon}$ are $2 \times2$ matrices and contain the spin-averaged part and spin-dependent part, i.e.,
\begin{eqnarray}
\hat{\varepsilon}(\vec{r},\vec{p},t)&=&\varepsilon(\vec{r},\vec{p},t)\hat{I}+\vec{h}(\vec{r},\vec{p},t)\cdot \vec{\sigma},\label{ener} \\
\hat{f}(\vec{r},\vec{p},t) &=&
f_{0}(\vec{r},\vec{p},t)\hat{I}+\vec{g}(\vec{r},\vec{p},t)\cdot\vec{\sigma},\label{fs}
\label{dens}
\end{eqnarray}
with $\vec{\sigma}=(\sigma_{x},\sigma_{y},\sigma_{z})$ being the Pauli matrices.

The spin-averaged and spin-dependent part of $\hat{\varepsilon}$ can be formally written as
\begin{eqnarray}
\varepsilon &=& \frac{p^2}{2m} + U_{MID} + \varepsilon_{so}, \\
\vec{h} &=& -\frac{W_{0}}{2}\nabla \times (\vec{j}+\vec{j}_{\tau}) +\frac{W_{0}}{2}\left [\nabla(\rho+\rho_{\tau}) \times \frac{\vec{p}}{\hbar}\right ],\label{h}
\end{eqnarray}
with
\begin{equation}
\varepsilon_{so} = -\frac{W_{0}}{2}\nabla\cdot (\vec{J}+\vec{J}_{\tau}) -\frac{W_{0}}{2}\left [\frac{\vec{p}}{\hbar} \cdot (\nabla\times (\vec{s}+\vec{s}_{\tau}))\right ]. \\
\end{equation}
In the above, $\tau=n,p$ is the isospin index,
\begin{equation}
U_{\rm MID} = a\left(\frac{\rho}{\rho_0}\right)+b\left(\frac{\rho}{\rho_0}\right)^c \pm 2E_{sym}^{pot}\left(\frac{\rho}{\rho_0}\right)^{\gamma_{sym}} \left(\frac{\rho_n-\rho_p}{\rho}\right) \label{umid}
\end{equation}
represents the momentum-independent mean-field potential, where ``$+(-)$" corresponds to the sign of the symmetry potential for neutrons (protons), $\rho_0=0.16$ fm$^{-3}$ is the saturation density, and parameters $a=-209.2$ MeV, $b=156.4$ MeV, $c=1.35$, $E_{sym}^{pot}=18$ MeV, $\gamma_{sym}=2/3$ are fitted by empirical properties of isospin asymmetric nuclear matter. $\varepsilon_{so}$ and $\vec{h}$ are obtained based on the Hartree-Fock approach~\cite{Engel:1975zz} from the Skyrme-type spin-orbit interaction~\cite{Vautherin:1971aw}
\begin{eqnarray}\label{vsoi}
v_{so} = i W_0 (\vec{\sigma}_1+\vec{\sigma}_2) \cdot \vec{k}^\prime \times
\delta(\vec{r}_1-\vec{r}_2) \vec{k},
\end{eqnarray}
where the spin-orbit coupling coefficient $W_0$ empirically ranges from 80 to 150 MeVfm$^5$ by studying shell structure in finite nuclei~\cite{Lesinski:2007ys,Zalewski:2008is,Bender:2009ty} and referencing various Skyrme forces~\cite{PhysRevC.85.035201,PhysRevC.82.024321}, $\vec{\sigma}_{1(2)}$ represents the Pauli matrices, $\vec{k}=(\vec{p}_1-\vec{p}_2)/2$ is the relative momentum operator, and $\vec{k}^\prime$ is the complex conjugate of $\vec{k}$. $\rho=\sum_\tau \rho_\tau$, $\vec{s}=\sum_\tau \vec{s}_\tau$, $\vec{j}=\sum_\tau \vec{j}_\tau$, and $\vec{J}=\sum_\tau \vec{J}_\tau$ are respectively the nucleon number density, the spin density, the kinetic density, and the spin-current density. In the semiclassical case and neglecting the isospin index, these densities can be expressed in terms of the spin-averaged and spin-dependent phase-space distribution functions as
\begin{eqnarray}
\rho(\vec{r}) &=& 2 \int d^{3}p f_0(\vec{r},\vec{p}), \label{rhor}\\
\vec{s}(\vec{r}) &=& 2 \int d^{3}p \vec{g}(\vec{r},\vec{p}),  \\
\vec{j}(\vec{r}) &=& 2 \int d^{3}p \frac{\vec{p}}{\hbar}f_0(\vec{r},\vec{p}), \label{rhoj} \\
\vec{J}(\vec{r}) &=& 2 \int d^{3}p \frac{\vec{p}}{\hbar} \times
\vec{g}(\vec{r},\vec{p}). \label{Jr}
\end{eqnarray}
In real simulations, these densities are calculated based on the test-particle method~\cite{Wong:1982zzb,Bertsch:1988ik} using the lattice-Hamiltonian approach~\cite{Lenk:1989zz}.

In the semiclassical case, a unit vector in the $4\pi$ solid angle is used to represent the expectation values of the Pauli matrices and thus the nucleon spin. The left-hand side of Eq.~(\ref{BLE}) can be solved in the leading-order by simulating the system evolution based on the following equations of motion~\cite{Xia:2016xiw}
\begin{eqnarray}
\frac{d\vec{r}_i}{dt} &=& \frac{\partial }{\partial \vec{p}_i} (\varepsilon + \vec{h} \cdot \vec{\sigma}_i),  \label{eom1}\\
\frac{d\vec{p}_i}{dt} &=& - \frac{\partial }{\partial \vec{r}_i} (\varepsilon + \vec{h} \cdot \vec{\sigma}_i), \label{eom2}\\
\frac{d\vec{\sigma}_i}{dt} &=& 2 \frac{\vec{h} \times \vec{\sigma}_i}{\hbar}, \label{eom3}
\end{eqnarray}
where $\vec{r}_i$, $\vec{p}_i$, and $\vec{\sigma}_i$ represents the coordinate, momentum, and spin of the $i$th particle. A stronger spin-orbit coupling from a larger $W_0$ enhances the difference in mean-field evolutions of nucleons with different spins [Eqs.~(\ref{eom1}) and (\ref{eom2})] and meanwhile a stronger spin precession [Eq.~(\ref{eom3})]. The simulation starts with nucleon coordinates in colliding nuclei sampled according to the spherical density distribution obtained from Skyrme-Hartree-Fock calculations, momenta sampled in the local Fermi sphere then boosted by the beam energy, and spins sampled randomly within the $4\pi$ solid angle. Then each nucleon evolves according to Eqs.~(\ref{eom1}), (\ref{eom2}), and (\ref{eom3}), with various densities in Eqs.~(\ref{rhor}-\ref{Jr}) updated at every time step based on the nucleon phase-space distribution evaluated by counting nucleon numbers and spins at each spatial cell.

For the collision term $I_c$ in Eq.~(\ref{BLE}), energy- and spin-dependent differential cross sections with certain angular distributions~\cite{Xia:2017dbx} obtained from the phase-shift data extracted from proton-proton and proton-neutron scatterings~\cite{PhysRevC.15.1002} are used. Spin- and isospin-dependent Pauli blockings are also implemented with more specified phase-space cells.

%\subsection{spin thermalized approach}

While the detailed spin dynamics in intermediate-energy heavy-ion collisions can be simulated based on the SIBUU model described above, the spin polarization from the spin-thermalized assumption will be compared to that from transport model calculations, with the vorticity and temperature fields extracted from the bulk properties of the nuclear matter dominated by spin-averaged dynamics. The non-relativistic vorticity field~\cite{Jiang:2016woz} can be calculated from
\begin{equation}\label{omeganr}
\vec{\omega}_{\rm NR} = \frac{1}{2} (\nabla \times \vec{v}),
\end{equation}
where $\vec{v}=\sum_i \vec{p}_i/\sum_i \sqrt{m^2+p_i^2}$ is the local velocity field with the summation over all nucleons within the local cell. In the non-relativistic limit and with the spin-thermalized assumption, the polarization of spin-$1/2$ particles is related to the vorticity field through~\cite{PhysRevC.95.054902}
\begin{equation}\label{P1}
\vec{P} = \frac{\vec{\omega}}{2T},
\end{equation}
where $T$ is the local temperature. Various methods can be used to extract the temperature of nuclear matter in heavy-ion collisions from transport simulations~\cite{PhysRevC.85.017604}. In SIBUU simulations, the number density $\rho_\tau$ of neutrons and protons as well as the kinetic energy density $\epsilon_k$ in the local rest frame are used to calculate inversely the local temperature $T$ and chemical potential $\mu_\tau$, i.e.,
\begin{eqnarray}
\rho_\tau &=& 2 \int \frac{d^3p}{(2\pi)^3} n_\tau, \\
\epsilon_k &=& 2 \sum_\tau \int \frac{d^3p}{(2\pi)^3} \frac{p^2}{2m} n_\tau,
\end{eqnarray}
where
\begin{equation}\label{nf}
n_\tau = \frac{1}{\exp(\frac{p^2/2m + U_{\rm MID}-\mu_\tau}{T}) + 1}
\end{equation}
is the local occupation probability.

The above kinematic vorticity has the following covariant form
\begin{equation}\label{omegaco}
\omega^\mu = - \frac{1}{2} \epsilon^{\mu\nu\rho\sigma} u_\nu \omega_{\rho\sigma},
\end{equation}
where $ \epsilon^{\mu\nu\rho\sigma}$ is the levi-civita symbol, $u_\mu = \gamma (1, -\vec{v})$ is the four-velocity field with $\gamma=1/\sqrt{1-v^2}$, and
\begin{equation}
\omega_{\mu\nu} = \frac{1}{2}(\partial_\nu u_\mu - \partial_\mu u_\nu)
\end{equation}
is the kinetic vorticity tensor. The thermal vorticity proposed in Ref.~\cite{Becattini:2013fla} has the form
\begin{equation}
\varpi^\mu = - \frac{1}{2} \epsilon^{\mu\nu\rho\sigma} u_\nu \varpi_{\rho\sigma},
\end{equation}
where
\begin{equation}
\varpi_{\mu\nu} = \frac{1}{2}(\partial_\nu \beta_\mu - \partial_\mu \beta_\nu)
\end{equation}
is the thermal vorticity tensor, with $\beta_\mu = u_\mu/T$. In the spin-vector approach~\cite{Becattini:2013fla,PhysRevC.94.024904,PhysRevC.95.054902}, the polarization of spin-$1/2$ particles can be obtained from
\begin{equation}\label{P2}
\vec{P}=2 \vec{S}^\star,
\end{equation}
where $\vec{S}^\star$ is the spin vector in the local rest frame, related to $\vec{S}$ in the calculational frame through
\begin{equation}
\vec{S}^\star = \vec{S} - \frac{\vec{p} \cdot \vec{S}}{p_0 (p_0 + m)} \vec{p}.
\end{equation}
In the above, $p_0=\sqrt{p^2+m^2}$ is the particle energy, and
\begin{equation}\label{ss}
S^\mu = -\frac{1}{8m} (1-n_\tau) \epsilon^{\mu\nu\rho\sigma} p_\nu \varpi_{\rho\sigma}
\end{equation}
represents the four-component spin vector, with $n_\tau$ calculated from Eq.~(\ref{nf}).

%\section{Results and discussions}
%\label{results}

Let's move to the comparison on the space-time evolution of the nucleon spin polarization during heavy-ion evolution and at the freeze-out stage from transport simulations and based on the spin-thermalized assumption. In SIBUU simulations, the coordinate, momentum, and spin of each nucleon can be traced, and the densities, phase-space distribution functions, and the resulting local spin polarization can be calculated during the non-equilibrium dynamic process. Empirically, nucleons freeze out at the local density lower than $\rho_0$/8 and become free ones that will no longer experience further scatterings or mean-field interactions. Based on the spin-thermalized assumption, the nucleon spin polarization is calculated from Eqs.~(\ref{P1}) or (\ref{P2}) using various local vorticity fields calculated based on the nucleon information provided by SIBUU simulations, and the spin polarization of final nucleons can be calculated at the location and time where/when they freeze out. Au+Au collisions at the beam energy of 100 AMeV and an impact parameter $\text{b}=8$ fm are used as the representative reaction for the illustration, with a default spin-orbit coupling coefficient $W_0=150$ MeVfm$^5$ used in SIBUU simulations stated otherwise.

\begin{figure}[ht]
\includegraphics[width=1\linewidth]{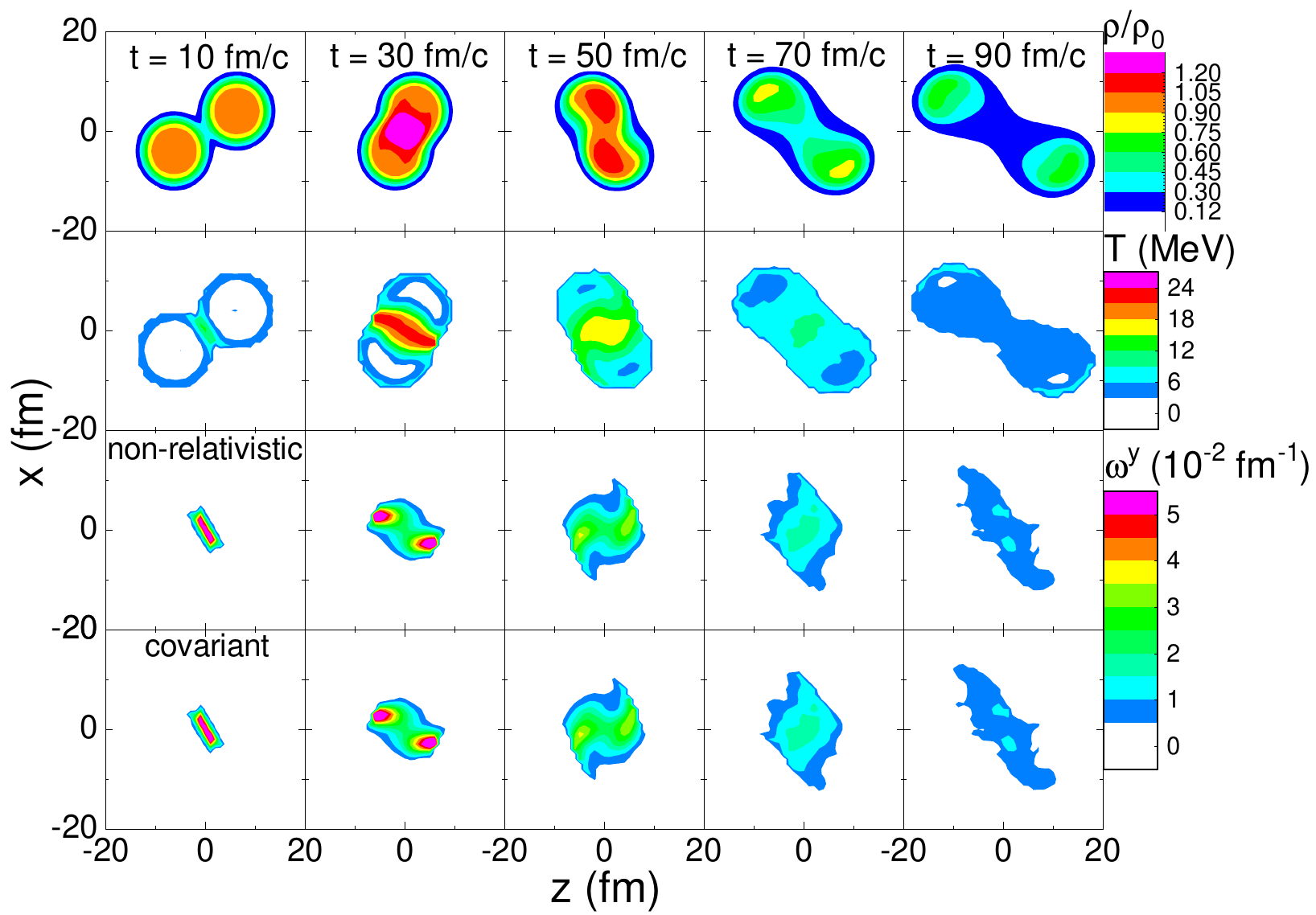}
\caption{\label{fig1} Reduced density $\rho/\rho_0$ (first row), temperature $T$ (second row), and the $y$-component of the non-relativistic (third row) and covariant (fourth row) kinematic vorticity in the reaction ($x-o-z$) plane at different times in non-central Au+Au collisions at the beam energy of 100 AMeV.}
\end{figure}

We begin by displaying various contours at different times in the reaction plane in Fig.~\ref{fig1}. The first row showing the density evolution provides a global picture of the reaction process, with the maximum density reached at about $t=30$ fm/c, and two nuclei do not interact after about $t=100$ fm/c. The temperature distribution is rather non-uniform at early stages, with $T$ higher than 20 MeV in the central compressed region, but becomes generally uniform as the system cools down at later stages, as shown in the second row. The evolutions of $y$-components of the kinematic vorticity calculated from Eqs.~(\ref{omeganr}) and (\ref{omegaco}) are displayed in the third and fourth row, respectively, and they have negligible difference, showing that the relativistic effect is unimportant at this collision energy. The kinematic vorticity reaches around $0.05$ fm$^{-1}$ in the most compressed stage, and becomes weaker at later stages.

\begin{figure}[ht]
\includegraphics[width=1\linewidth]{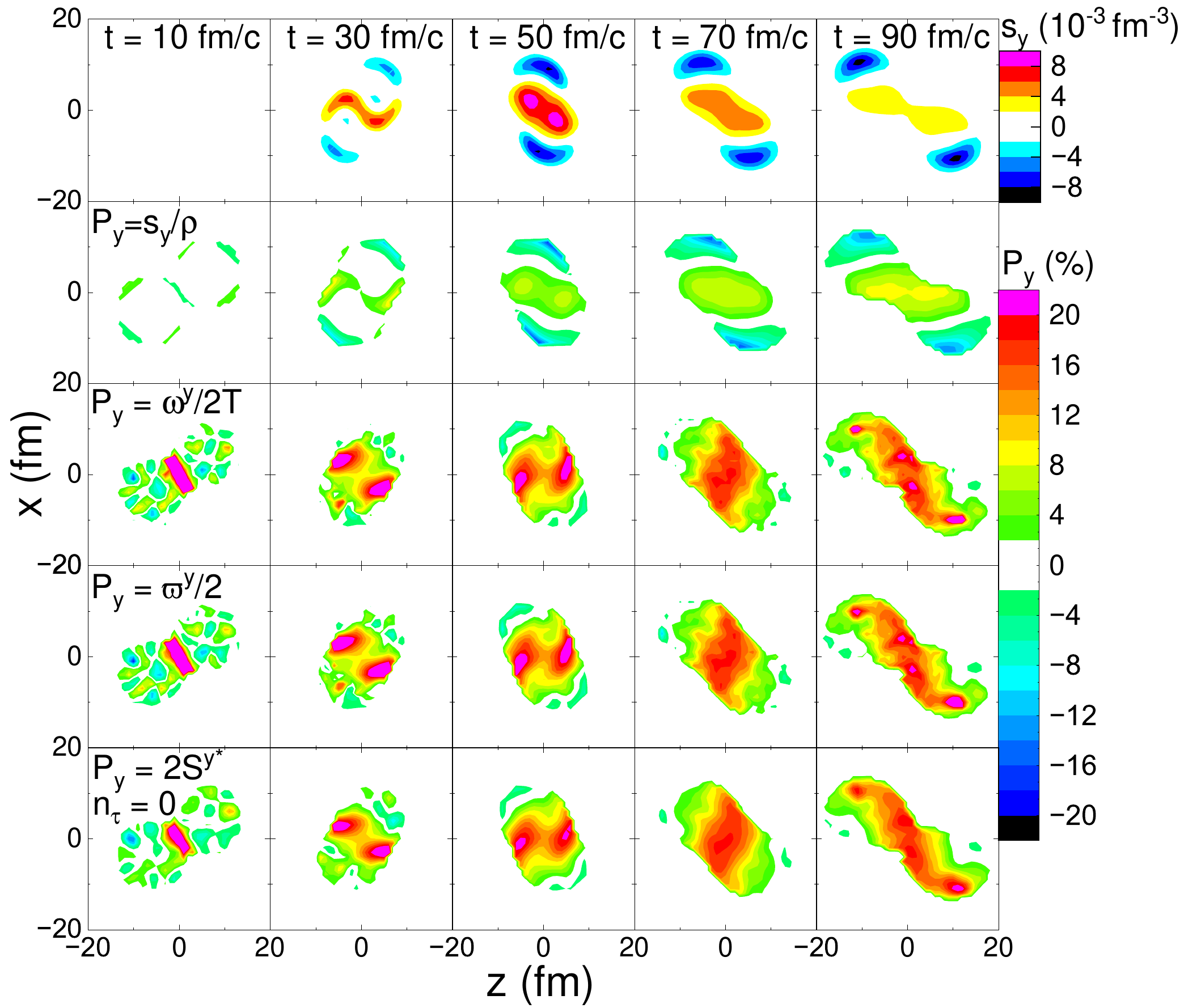}
\caption{\label{fig2} $y$-component of the spin density $s_y$ (first row) and the corresponding nucleon spin polarization $P_y=s_y/\rho$ (second row) in $y$ direction from SIBUU, as well as the nucleon spin polarization $P_y$ from the kinematic vorticity (third row), the thermal vorticity (fourth row), and the spin vector (fifth row) in the reaction plane at different times in non-central Au+Au collisions at the beam energy of 100 AMeV.}
\end{figure}

Figure~\ref{fig2} compares the evolution of the spin polarization perpendicular to the reaction plane from different approaches. The first row shows the $y$-component of the spin density $s_y$ calculated from SIBUU, and the spin polarization in the second row is simply calculated from $s_y/\rho$. The spin polarization in SIBUU is generated by the spin-orbit potential as Eq.~(\ref{h}). While the two terms in Eq.~(\ref{h}) cancel each other in a free-moving nucleus, the first term dominates the dynamics during the interaction stage. This attracts nucleons with spin $+\hat{y}$ to the central participant region and repels those with spin $-\hat{y}$ to the spectator region, leading to a positive $P_y$ in the participant matter but a negative $P_y$ in the spectator matter. In the third, fourth, and fifth row, the nucleon spin polarization is calculated based on $P_y = \omega^y/2T$ from the kinematic vorticity, $P_y = \varpi^y/2$ from the thermal vorticity, and $P_y = 2 {S^y}^\star$ from the spin vector without the Pauli blocking factor, respectively. Only small differences from different approaches are observed, showing that the temperature gradient is largely cancelled by the corresponding covariant term and thus does not affect much the nucleon spin polarization, different from the situation in relativistic heavy-ion collisions where both the velocity and temperature fields change rapidly with space and time. On the other hand, the spin-thermalized approaches do not predict the negative polarization for spectator matter, but give a positive polarization as high as $20\%$ in the participant region compared to that of about $8\%$ from SIBUU in the late stage of the collision.

\begin{figure}[ht]
\includegraphics[width=1\linewidth]{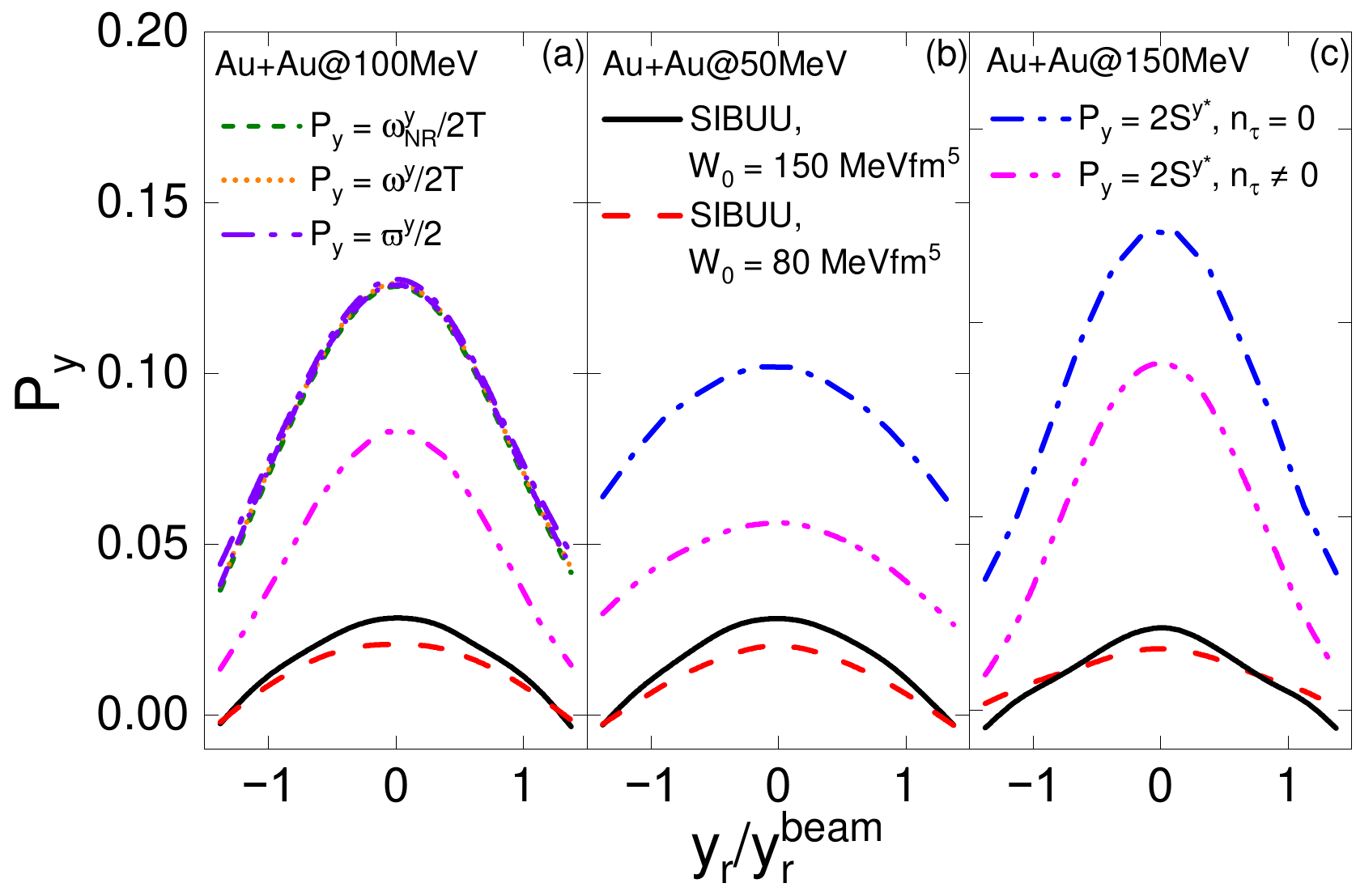}
\caption{\label{fig3} Rapidity dependence of the spin polarization $P_y$ for free nucleons in non-central Au+Au collisions at the beam energy of 100 (a), 50 (b), and 150 (c) AMeV, from SIBUU with different spin-orbit coupling coefficients $W_0$, and various approaches based on the spin-thermalized assumption.}
\end{figure}

Figure~\ref{fig3} compares the rapidity dependence of the spin polarization $P_y$ perpendicular to the reaction plane for free nucleons from various approaches and different collision energies. The magnitudes of $P_y$ for free nucleons are mostly determined by those in the low density regions at later stages in the collision as shown in Fig.~\ref{fig2}. While $P_y$ is generally larger at midrapidity but smaller at large rapidities in all cases, the approaches with the spin-thermalized assumption predict a $P_y$ as large as $\sim 10\%$ compared with a $P_y$ of a few percent from SIBUU simulations depending on the spin-orbit coupling coefficient. All spin-thermalized approaches give similar results, except that the $P_y$ is considerably reduced after the Pauli blocking factor $1-n_\tau$ in Eq.~(\ref{ss}) is taken into account. The spin-thermalized approaches predict that $P_y$ generally increases with increasing collision energy due to the stronger vorticity field as well as the weaker Pauli blocking at higher collision energies. SIBUU predicts a maximum $P_y$ around $50 \sim 100$ AMeV but smaller $P_y$ at higher collision energies due to the stronger spin procession at higher collision energies according to Eq.~(\ref{eom3}).

\begin{figure}[ht]
\includegraphics[width=1\linewidth]{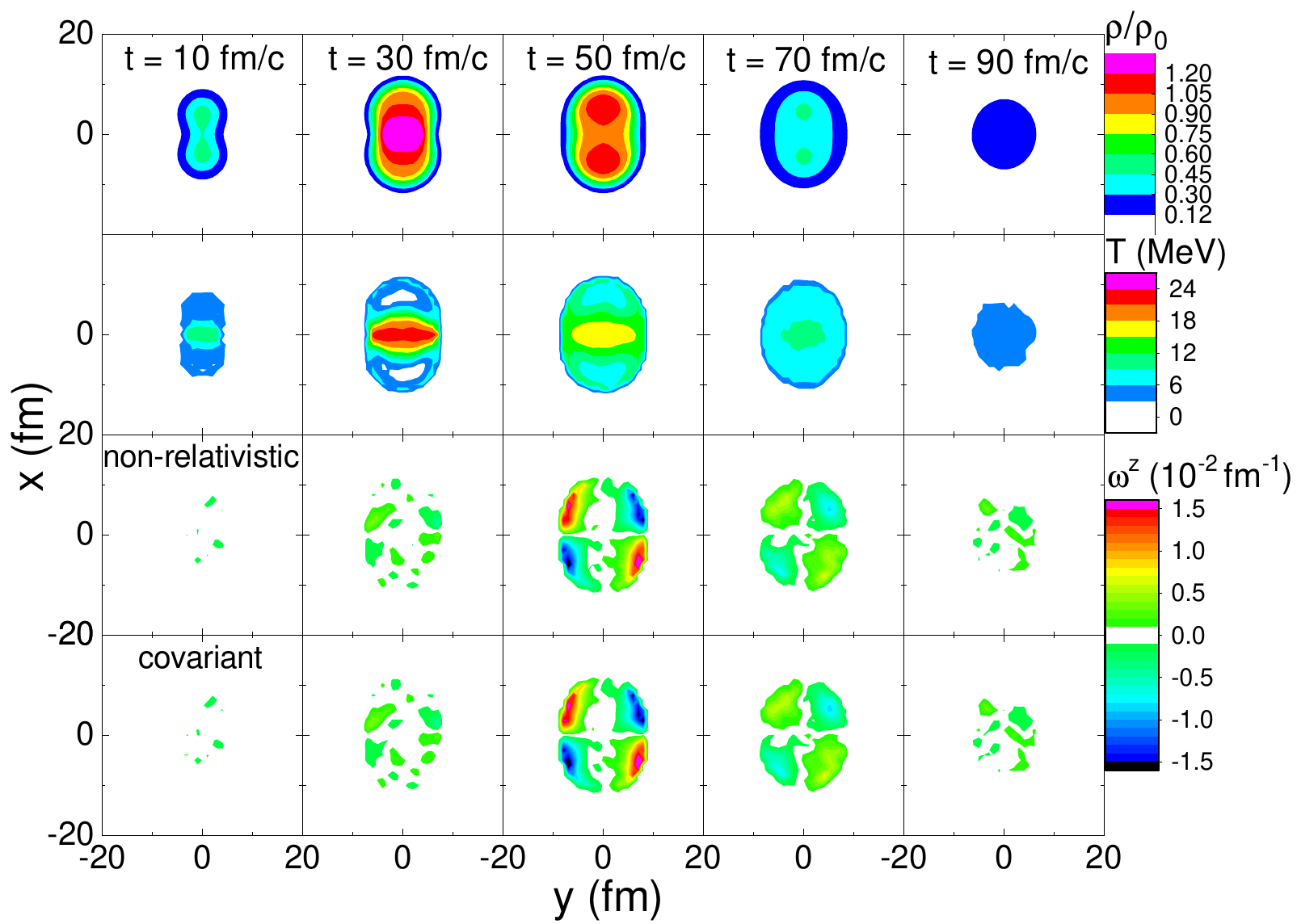}
\caption{\label{fig4} Reduced density $\rho/\rho_0$ (first row), temperature $T$ (second row), and $z$-component of the non-relativistic (third row) and covariant (fourth row) kinematic vorticity in the transverse ($x-o-y$) plane at different times in non-central Au+Au collisions at the beam energy of 100 AMeV.}
\end{figure}

\begin{figure}[ht]
\includegraphics[width=1\linewidth]{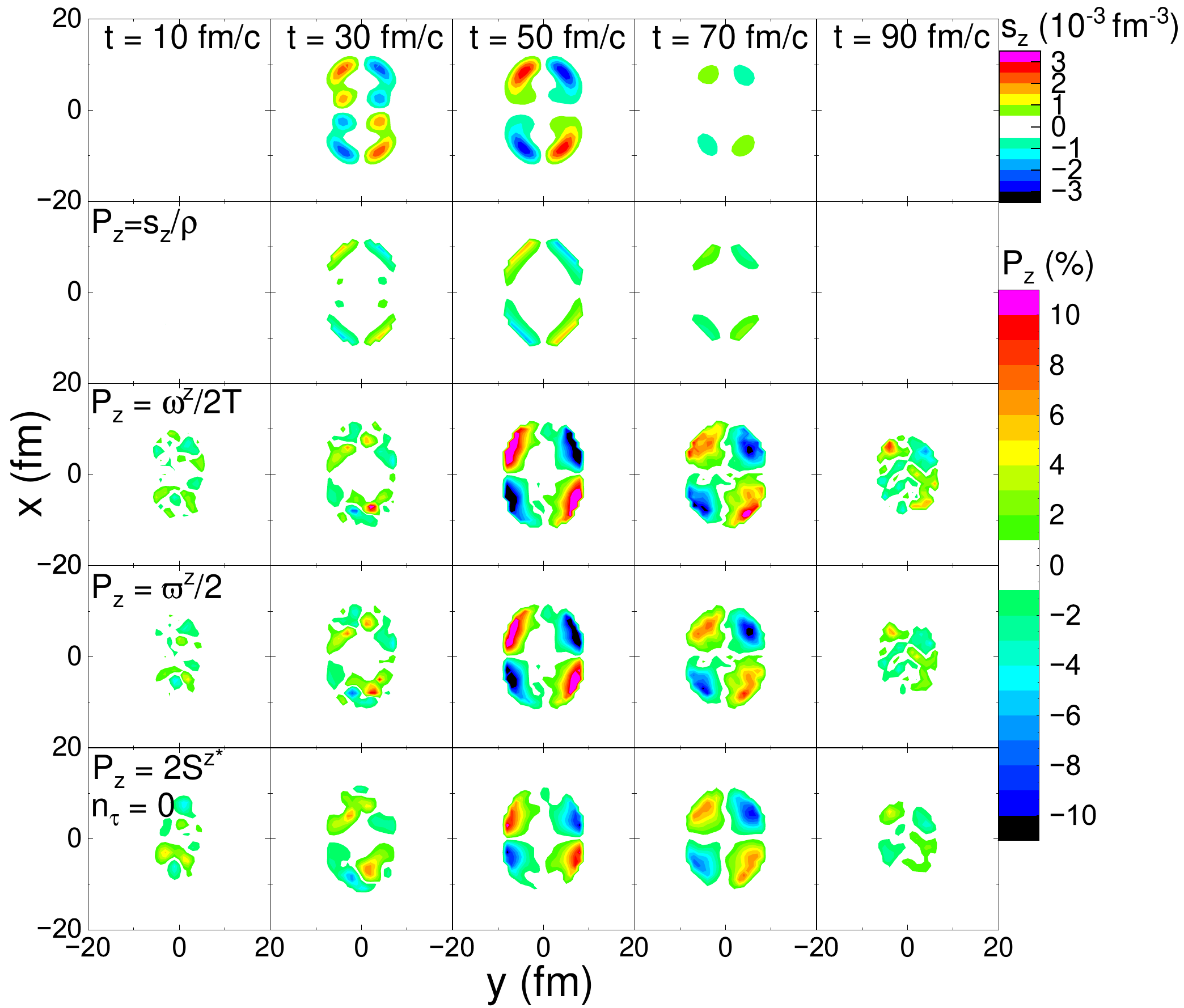}
\caption{\label{fig5} $z$-component of the spin density $s_z$ (first row) and the corresponding nucleon spin polarization $P_z=s_z/\rho$ (second row) in $z$ direction from SIBUU, as well as the nucleon spin polarization $P_z$ from the kinematic vorticity (third row), the thermal vorticity (fourth row), and the spin vector (fifth row) in the reaction plane at different times in non-central Au+Au collisions at the beam energy of 100 AMeV.}
\end{figure}

\begin{figure}[ht]
\includegraphics[width=1\linewidth]{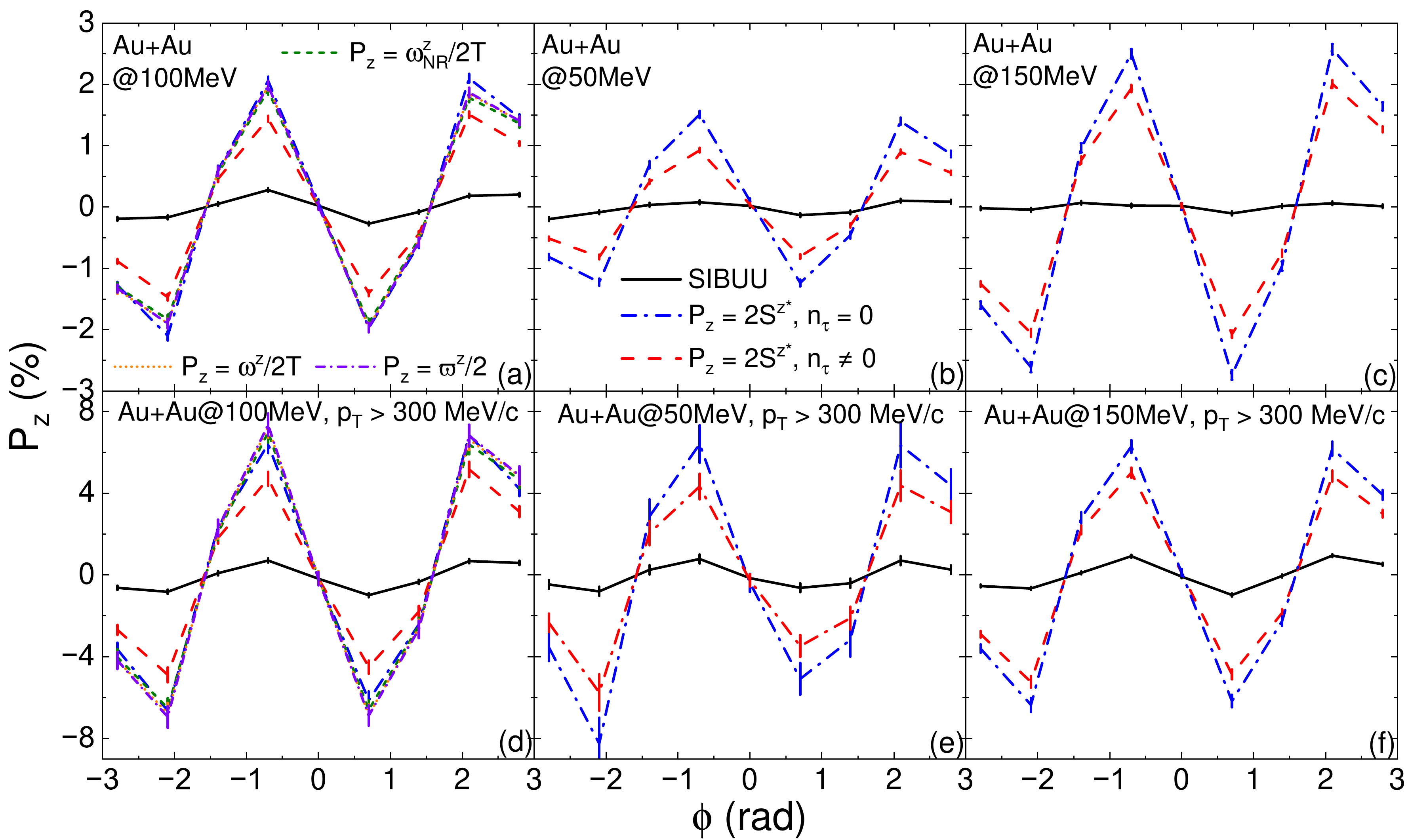}
\caption{\label{fig6} Azimuthal angle dependence of the spin polarization $P_z$ for mid-rapidity ($|y_r|<y_r^{\rm beam}/2$) free nucleons in non-central Au+Au collisions at the beam energy of 100 (left), 50 (middle), and 150 (right) AMeV, from SIBUU and various approaches based on the spin-thermalized assumption. Results for all free nucleons are shown in upper panels, and those for energetic free nucleons ($p^{}_T>300$ MeV/c) are shown in low panels.}
\end{figure}

Now let's move to the comparison on the longitudinal spin polarization. The first and second row in Fig.~\ref{fig4} show similar time evolutions of the reduced density and the temperature distributions as in Fig.~\ref{fig1} but in the transverse plane. The third and fourth row in Fig.~\ref{fig4} show the $z$-component of the kinematic vorticity from non-relativistic and covariant calculations, and they are very similar. One sees that the maximum density and the highest temperature are reached around $t=30$ fm/c, while the strongest vorticity field is formed around $t=50$ fm/c. Different from the positive $\omega^y$ in the reaction plane, $\omega^z$ can be positive or negative and shows some spatial structures. Figure~\ref{fig5} compares the time evolutions of the local spin polarization from various approaches. In SIBUU simulations, the non-zero $z$-component of the spin density $s_z$ is again generated by the spin-orbit potential [Eq.~(\ref{h})] as shown in the first row, and the second row shows the corresponding longitudinal spin polarization $P_z=s_z/\rho$. The third, fourth, and fifth row compare time evolutions of the longitudinal polarization from the kinematic vorticity $P_z = \omega^z/2T$, the thermal vorticity $P_z = \varpi^z/2$, and the spin vector $P_z = 2 {S^z}^\star$ without the Pauli blocking factor, respectively. Results based on the spin-thermalized assumption are similar, and they can be as large as $\pm 10 \%$ compared to $\pm 4 \%$ from SIBUU around $t=50$ fm/c. $P_z$ from all approaches become weaker at later stages.

Figure~\ref{fig6} compares the azimuthal angle dependence of the longitudinal spin polarization for mid-rapidity free nucleons from various approaches and at different collision energies. The upper panels show results from all free nucleons, while the lower panels show results from energetic free nucleons that most emit at early stages corresponding to a stronger longitudinal spin polarization as shown in Fig.~\ref{fig5}. The dependence of $P_z$ on the azimuthal angle $\phi = {\rm atan2}(p_y,p_x)$ for free nucleons are all consistent with the spatial distribution of $P_z$ shown in Fig.~\ref{fig5}, with a positive $P_z$ around $\phi=-\pi/3$ and $\phi=2\pi/3$ and a negative $P_z$ around $\phi=-2\pi/3$ and $\phi=\pi/3$. Again, the spin-thermalized approaches predict much stronger longitudinal spin polarization than SIBUU, even with a reduction by the Pauli blocking factor in the spin-vector approach. The $P_z$ from all free nucleons is enhanced with increasing collision energy from spin-thermalized approaches, but does not show a monotonical dependence on the collision energy from SIBUU.

%\section{Conclusion and outlook}
%\label{summary}

To summarize, nucleon spin polarizations during the collision process and at the freeze-out stage from transport model simulations and the spin-thermalized assumption in intermediate-energy heavy-ion collisions have been compared. In the collision energy considered in the present study, the relativistic effect and the temperature gradient are unimportant but the Pauli blocking factor reduces the spin polarization considerably in the spin-thermalized assumption. All spin-thermalized approaches overestimate significantly the spin polarization perpendicular to the reaction plane and in the longitudinal direction, compared with those generated by the spin-orbit mean-field potential in transport model simulations.

While there is still no experimental data of the nucleon spin polarization in heavy-ion collisions so far, it has been proposed to measure proton spin polarizations using, e.g., $^{12}$C with known analyzing power as an effective detector~\cite{c642-1lzb}, which may hopefully confirm the prediction by the present study. On the theoretical side, the spin-orbit potential is the main driving force to the spin dynamics in intermediate-energy heavy-ion collisions, while the tensor force has much weaker effects~\cite{Xu:2015kxa}. The spin change after nucleon-nucleon collisions~\cite{Liu:2025vho}, even with the in-medium correction, is not likely to fill the gap between results from SIBUU and the spin-thermalized assumption observed in Figs.~\ref{fig3} and \ref{fig6}. At higher collision energies where the relativistic effects become important, a covariant framework is needed to describe the spin dynamics.

%\begin{acknowledgments}
This work is supported by the National Natural Science Foundation of China under Grant Nos. 12375125, 11922514, and 11475243, and the Fundamental Research Funds for the Central Universities.
%\end{acknowledgments}

\bibliography{sibuu}
\end{document}